\documentclass[twocolumn]{aastex631}

\graphicspath{{./}{figures/}}
\usepackage{amsmath}

\usepackage{tensor}
\usepackage{multirow}
\usepackage{makecell}

\usepackage{CJK}

\usepackage[T1]{fontenc}
\usepackage{xspace}

\defcitealias{Narayan2022}{RN22}
\newcommand{\rn}{\citetalias{Narayan2022}\xspace}

\newcommand{\bfluxc}{\texttt{bflux-const}}
\newcommand{\R}{\mathcal{R}}

\makeatletter
\newcommand\approxsim{\mathpalette\@approxsim\relax}
\newcommand\@approxsim[2]{%
  \mathrel{%
    \ooalign{%
      $\m@th#1\sim$\cr
      \hidewidth$\m@th#1:$\hidewidth\cr
    }%
  }%
}
\makeatother

\newcommand\Tstrut{\rule{0pt}{3ex}}

\newcommand{\appropto}{\mathrel{\vcenter{
  \offinterlineskip\halign{\hfil$##$\cr
    \propto\cr\noalign{\kern2pt}\sim\cr\noalign{\kern-2pt}}}}}

\shorttitle{BH Subgrid Prescription from First-Principles GRMHD}
\shortauthors{Cho et al.}

\begin{document}
\begin{CJK}{UTF8}{mj}
\title{Bridging Scales in Black Hole Accretion and Feedback: Subgrid Prescription from First Principles}

\author[0000-0002-2858-9481]{Hyerin Cho (조혜린)}
\affiliation{Center for Astrophysics $\vert$ Harvard \& Smithsonian, 60 Garden Street, Cambridge, MA 02138, USA}
\affiliation{Black Hole Initiative at Harvard University, 20 Garden Street, Cambridge, MA 02138, USA}

\author[0000-0002-0393-7734]{Ben S. Prather} 
\affiliation{Center for Astrophysics $\vert$ Harvard \& Smithsonian, 60 Garden Street, Cambridge, MA 02138, USA}
\affiliation{Black Hole Initiative at Harvard University, 20 Garden Street, Cambridge, MA 02138, USA}

\author[0000-0002-1919-2730]{Ramesh Narayan}
\affiliation{Center for Astrophysics $\vert$ Harvard \& Smithsonian, 60 Garden Street, Cambridge, MA 02138, USA}
\affiliation{Black Hole Initiative at Harvard University, 20 Garden Street, Cambridge, MA 02138, USA}

\author[0000-0003-1598-0083]{Kung-Yi Su}
\affiliation{Department of Physics \& Astronomy and Center for Interdisciplinary Exploration and Research in Astrophysics(CIERA), Northwestern University, 1800 Sherman Ave, Evanston, IL 60201, USA}
\affiliation{Center for Astrophysics $\vert$ Harvard \& Smithsonian, 60 Garden Street, Cambridge, MA 02138, USA}
\affiliation{Black Hole Initiative at Harvard University, 20 Garden Street, Cambridge, MA 02138, USA}

\author[0000-0001-5287-0452]{Angelo Ricarte}
\affiliation{Center for Astrophysics $\vert$ Harvard \& Smithsonian, 60 Garden Street, Cambridge, MA 02138, USA}
\affiliation{Black Hole Initiative at Harvard University, 20 Garden Street, Cambridge, MA 02138, USA}

\author[0000-0002-5554-8896]{Priyamvada Natarajan}
\affiliation{Black Hole Initiative at Harvard University, 20 Garden Street, Cambridge, MA 02138, USA}
\affiliation{Department of Astronomy, Yale University, Kline Tower, 266 Whitney Avenue, New Haven, CT 06511, USA}
\affiliation{Department of Physics, Yale University, P.O. Box 208121, New Haven, CT 06520, USA}

\author[0000-0002-1996-0445]{Antonio J. Porras-Valverde}
\affiliation{Department of Astronomy, Yale University, Kline Tower, 266 Whitney Avenue, New Haven, CT 06511, USA}

\begin{abstract} \label{abstract}
Understanding how supermassive black holes (BHs) couple to their host galaxies across a vast spatial and temporal dynamic range remains a central challenge in galaxy evolution. Using the multizone framework---designed to capture bidirectional inflow--outflow from the event horizon to the Bondi scale---we present a suite of long-duration GRMHD simulations spanning BH spins $|a_\ast|=0$--0.9 and Bondi radii $R_B/r_g=4\times10^2$--$2\times10^6$. From these simulations we derive spin-dependent subgrid prescriptions from first principles, applicable to hot accretion flows with low-Eddington ratios ($f_{\rm Edd}\lesssim10^{-3}$), for adoption in cosmological simulations and semi-analytic models. We provide compact analytic fits for the time-averaged accretion rate $\dot M(R_B,a_\ast)$ and feedback power $\dot E_{\rm fb}(R_B,a_\ast)$ with respect to the Bondi rate $\dot{M}_B$, which are largely insensitive to the initial gas configuration and magnetic field strength. To capture intrinsic time-variability, we also quantify the full distributions of $\dot M$ and feedback efficiency $\eta$, both well described by lognormal statistics, with widths that increase toward larger $R_B$. We further measure self-consistent spin evolution in the hot accretion mode, finding that the spinup parameter varies as $s(a_\ast)\simeq -3.7\,a_\ast$, which implies a very long spindown timescale $t_s\simeq 12(10^{-3}/f_{\rm Edd})\,{\rm Gyr}$. Thus, BH spins are effectively frozen during phases of quiescent accretion. Compared to conventional small-domain GRMHD calculations, our simulations, which reach dynamical equilibrium across horizon-to-galaxy scales, yield systematically different long-term accretion, feedback, and spin properties, cautioning against direct extrapolation from small-scale GRMHD simulations when constructing galactic-scale subgrid models.
\end{abstract}

\section{Introduction} \label{sec:intro}

Supermassive black holes (BHs) at the centers of massive galaxies are thought to co-evolve with their hosts \citep[see e.g.,][for reviews]{Natarajan2011,Kormendy2013,Heckman2014}. Galaxies supply gas for BH accretion, while BHs return energy and momentum via feedback, establishing an accretion--feedback loop that can quench star formation in massive ``red and dead'' systems. At low Eddington ratios, typical of many nearby supermassive BHs, accretion proceeds through hot, radiatively inefficient, advection dominated accretion flows (ADAFs; see \citealt[for a review]{Yuan2014}). In this regime, an important feedback channel is the \citet{Blandford1977} mechanism, in which magnetic fields threading a spinning BH launch relativistic jets that extract the BH's rotational energy, a magnetic-field analogue of the \cite{Penrose_69} process \citep{Komissarov2009,Lasota2014}.

The BH--galaxy coupling spans an enormous range of scales. In M87, a Schwarzschild radius of $\sim0.6\,{\rm mpc}$ connects to jet emission observed from tens of Schwarzschild radii \citep{Lu2023,EHTMWLScienceWorkingGroup2021} out to $\sim20\,{\rm kpc}$ radio lobes \citep{Owen2000}. Capturing such two-way coupling poses a major computational challenge. Horizon-scale GRMHD simulations \citep[e.g.,][]{Komissarov1999,Gammie2003,Sadowski2013, White2016,Porth:2019, Liska2022, Prather2024} self-consistently resolve accretion and jet launching but typically achieve steady states only out to $\lesssim$ a few hundred horizon radii. By contrast, cosmological and galaxy simulations \citep[e.g.,][]{Tremmel2017,Weinberger2018,Dave2019,Ni2022,Wellons2023,Schaye2025} require subgrid prescriptions for unresolved BH accretion and feedback. Zoom-in/hyper-refinement approaches, which begin with coarse resolution and progressively enhance it \citep[e.g.,][]{Hopkins2010,Ressler2020b,Angles-Alcazar2021,Guo2023,Guo2024,Hopkins2024,Hopkins2024b,Hopkins+2025ISCO,Kaaz2025}, can track inflow from galactic to BH scales, but still struggle to propagate feedback back to the galaxy.

A key uncertainty in cosmological simulations is therefore the BH subgrid model: differing implementations produce a wide range of predictions \citep[e.g.,][]{Habouzit2022,Haidar2022,Weinberger2023,Wellons2023}. Some physically-motivated approaches adopt highly idealized analytic prescriptions for accretion \citep[e.g.,][]{Bondi1952,Shakura1973,Novikov1973,Narayan1994} and feedback \citep[e.g.,][]{Blandford1977}, or incorporate spin-dependent feedback \citep{Dubois2021,Talbot2021,Beckmann2025,Husko2025} calibrated on small-domain GRMHD calculations \citep{Tchekhovskoy2012,McKinney2012,Ricarte2023,Lowell2024}. However, how reliably small-scale GRMHD results carry over to galactic scales remains largely untested.

The ``multizone method'' developed by our group bridges this gap, enabling first-principles GRMHD simulations that connect the horizon to beyond the Bondi radius across $\sim$8 orders of magnitude \citep{Cho2023,Cho2024,Cho2025,Su2025}. Subsequently, similar scale-bridging methods have since been implemented for Cartesian grids \citep{Guo2025} and for Lagrangian simulations \citep{Hopkins2025}.

For non-spinning BHs, \citet{Cho2023,Cho2024} showed that magnetic fields strongly suppress $\dot{M}$ relative to the Bondi prediction $\dot{M}_B$ \citep{Bondi1952}, with the suppression depending on the Bondi radius $R_B$ (defined in \autoref{eq:RB}). Long evolutions reach the magnetically arrested disk (MAD) state (\citealt{Narayan2003,Igumenshchev2003}, see also \citealt{Bisnovatyi-Kogan1974,Bisnovatyi1976}), where saturated magnetic flux drives highly non-axisymmetric inflow via interchange instabilities. Magnetic fields also strongly brake rotation, effectively erasing memory of the initial gas angular momentum \citep{Cho2024}. For spinning BHs ($a_\ast=0.9$), \citet{Cho2025} found a time-averaged feedback efficiency $\eta\approx 0.3$ (defined in \autoref{eq:eta_def}) independent of $R_B$ and regardless of the initial gas rotation state. The insensitivity of the final state to the initial gas rotation is consistent with \citet{Cho2025v}'s predictions that systems eventually evolve to a strongly variable state, characterized by flips in rotation and large jet-power fluctuations \citep[e.g.,][]{Ressler2021,Kwan2023,Lalakos2024, Cho2024,Cho2025,Galishnikova2025,Kim2025,Guo2025,Chan2025,Lalakos2025}. Some of the early multizone results have already begun to be incorporated into galaxy simulations \citep{Su2025}, cosmological simulations \citep{Su2025b}, and semi-analytic models (SAMs) \citep{Porras-Valverde2025}.

In this work, we extend the multizone method to a grid of BH spins, $a_\ast=0, ~0.1, ~0.3, ~0.5, ~0.7, ~0.9$, and explore the dependence on $R_B$ at each spin. From this library, we derive simple subgrid prescriptions for the time-averaged BH accretion rate $\dot{M}$ and feedback power $\dot{E}_{\rm fb}$ that are obtained directly from large-scale, first-principles GRMHD simulations. They are not tuned to observations, and are straightforward to implement in cosmological simulations and SAMs. We describe the numerical method in Section~\ref{sec:method}, present the subgrid prescriptions in Section~\ref{sec:subgrid}, outline spin evolution in Section~\ref{sec:spin_evolution}, and conclude in Section~\ref{sec:conclusion}.

Throughout this work, we adopt units where $GM=c=1$, so our units of length and time are the gravitational radius $r_g\equiv GM /c^2$ and the gravitational time $t_g\equiv r_g/c$. Although our GRMHD simulations are scale-free and are, in principle, applicable to any BH mass $M$, we focus on supermassive BHs in galactic centers. We define the Bondi radius as
\begin{equation}\label{eq:RB}
    R_B \equiv\frac{GM}{c_{s,\infty}^2} \approx \frac{7\times 10^{12}\,{\rm K}}{\gamma_{\rm ad}T_{\infty}}\,r_g,
\end{equation}
where $c_{s,\infty}$  and $T_\infty$ are the sound speed and temperature (in K) of the gas outside the Bondi radius. We use an adiabatic index $\gamma_{\rm ad} =5/3$. The Bondi timescale $t_B$ is defined as the free-fall timescale measured at the Bondi radius: $t_B\equiv (R_B/r_g)^{3/2}\,t_g$. 

The low Eddington-ratio accretion flows we study in this work typically consist of hot coronal gas from the galaxy flowing in at the Bondi radius and accreting on the supermassive BH. The temperature $T_\infty$ in \autoref{eq:RB} is then of order the virial temperature of the galaxy, which
is $T_\infty \sim 10^8\,{\rm K}$ in the case of large galaxy clusters \citep[e.g.,][]{Vikhlinin2005,Wallbank2022}; $T_\infty \gtrsim 10^7$\,K for giant elliptical galaxies like M87 \citep{DiMatteo2003,Russell2015}; $T_\infty \sim 10^6-10^7$\,K for the Milky Way \citep{Baganoff2003}; and $T_\infty \sim10^6$\,K for smaller galaxies \citep[e.g.,][]{Ott2005}.\footnote{\citet{Su2025b} find that $T_\infty$, which is the average temperature in the vicinity of BHs at the limited resolution of cosmological simulations, rarely exceeds $10^8\,{\rm K}$.} Therefore, from \autoref{eq:RB}, a realistic range of Bondi radii to consider for investigations is $R_B \sim ({\rm few}\times 10^4-{\rm few}\times 10^6)\,r_g$. 

\section{Numerical Method}\label{sec:method}
Our simulations utilize the GRMHD code KHARMA\footnote{https://github.com/AFD-Illinois/kharma} (``Kokkos-based High-Accuracy Relativistic Magnetohydrodynamics with Adaptive mesh refinement,'' \citealt{Prather2024}). Given a fixed Kerr metric $g_{\mu\nu}$, KHARMA solves the ideal GRMHD equations for the conserved quantities $\rho u^t$, $T^t_\mu$, and $B^i$. Here, $T^{\mu\nu}=\left( \rho + u + p_g + b^2 \right) u^{\mu} u^{\nu} + \left( p_g + b^2/2 \right) g^{\mu \nu} - b^{\mu}b^{\nu}$ is the energy-momentum tensor of the magnetized gas, $\rho$ is the rest mass density, $u$ is the internal energy density, $p_g=(\gamma_{\rm ad}-1)u$ is the gas pressure, $b^\mu$ is the magnetic field four-vector, and $u^\mu$ is the fluid four-velocity. We refer the reader to \citet{Anile1990,Komissarov1999,Gammie2003} for details of the GRMHD equations. As with most GRMHD simulations, radiative effects are ignored, which is a valid approximation for ADAFs with low accretion rates, $\dot{M}\lesssim 10^{-3}\dot{M}_{\rm Edd}$, where $\dot{M}_{\rm Edd}$ is the Eddington rate.

The multizone method significantly accelerates simulations of problems involving a vast range of length- and time-scales. This is achieved by sequentially iterating between small and large timesteps, allowing each scale to evolve on its own characteristic timescale. Conventional simulations, in contrast, are constrained to use timesteps set by the horizon, which is extraordinarily short compared to the timescale on which any meaningful large-scale evolution occurs. Also, the multizone method is designed to allow sufficient communication between small and large scales, helping the system to efficiently reach a global dynamical equilibrium on all scales. The method is described and tested in detail in \citet{Cho2024,Cho2025}, so here we provide only a brief overview, along with additional numerical details, in Appendix~\ref{sec:method_detail}.

All simulations, except those described in Appendix \ref{sec:dependence_ic}, use the same initial conditions (ICs), similar to a spherically symmetric Bondi flow (called {\bf B} IC in \citealt{Cho2025}). The density is initialized with a profile $\rho_{\rm init}(r)\propto (r+R_B)/r$ and the temperature $T$ and four-velocity $u^\mu$ are set to the general relativistic analytical solution of the hydrodynamic Bondi accretion problem \citep{Michel1972,Shapiro1983}. Magnetic fields are initially axisymmetric and nearly vertical, with plasma-$\beta\equiv 2\rho T/b^2$, or gas-to-magnetic pressure ratio, of order unity at all radii. There is no gas angular momentum to begin with. However, we note that using ICs with rotating gas or sub-dominant magnetic fields ($\beta\sim 100$) has a minor effect on the final state, as shown in \citet{Cho2025} and demonstrated again in Appendix~\ref{sec:dependence_ic}. The total duration of each simulation is $700\,t_B$ for all runs, unless otherwise stated.

\subsection{Diagnostics}

For the following analyses, the accretion rate is defined as 
\begin{equation}
\dot{M}(r) \equiv - \iint \rho u^r \sqrt{-g}\,d\theta\,d\varphi.
\end{equation}
We choose $\dot{M}(5\,r_g)$ as the nominal accretion rate of the BH.
The energy inflow rate is $\dot{E}(r) \equiv \iint T^r_t \sqrt{-g}\,d\theta\,d\varphi$.
The net feedback power is the difference between the energy outflow rate $-\dot{E}$ and the outflow rate of rest-mass energy $-\dot{M}$:
\begin{equation}
    \dot{E}_{\rm fb}(r) \equiv \dot{M}-\dot{E}.
\end{equation}
The dimensionless feedback efficiency is then
\begin{equation}\label{eq:eta_def}
    \eta(r) \equiv \dot{E}_{\rm fb}(r) / (\dot{M}(5\,r_g)c^2).
\end{equation}
The kinetic and thermal energy components of the total feedback efficiency are calculated as $\eta_{\rm kinetic}(r)\equiv (-\iint \rho u^r (u_t + 1)\sqrt{-g}\,d\theta\,d\varphi)/(\dot{M}(5\,r_g)c^2)$ and $\eta_{\rm thermal}(r)\equiv (-\iint (u+p_g) u^r u_t \sqrt{-g}\,d\theta\,d\varphi)/(\dot{M}(5\,r_g)c^2)$, respectively. 
To estimate the BH spin evolution, we need the angular momentum inflow flux, which is defined as
\begin{equation}\label{eq:Jdot}
    \dot{J}(r) = -\iint T^r_\varphi\sqrt{-g}\,d\theta\,d\varphi.
\end{equation}
Time averages of quantities are indicated with an overbar, e.g., $\overline{\dot{M}}$, $\overline{\dot{E}_{\rm fb}}$, $\overline{\eta}$. If the radius is not specified, the accretion rate $\dot{M}$ is measured near the horizon at $5\,r_g$ and the feedback power $\dot{E}_{\rm fb}$ and efficiency $\eta$ are measured near the Bondi radius at $R_B/3$.\footnote{Following \citet{Cho2025} and out of an abundance of caution, the BH feedback is measured just inside the Bondi radius (at $R_B/3$) because the gas dynamics undergoes a large transition across $R_B$. In practice, the feedback efficiency $\eta(r)$ remains quite constant as a function of radius, as demonstrated in \autoref{fig:eta_profiles}, so the main results are not impacted by where the measurements are made.} The BH mass accretion rate $\dot{M}$ is usually expressed in units of the classic Bondi accretion rate, which for $\gamma_{\rm ad}=5/3$ is
\begin{equation}\label{eq:MdotB} 
\dot{M}_B= \pi G^2 M^2 \overline{\rho(R_B)}/c^3_{s,\infty}.
\end{equation}
When estimating $\dot{M}_B$, we use the time-averaged gas density, $\overline{\rho(R_B)}$, measured at $R_B$ in the simulation (as in \citealt{Cho2024}).

\section{Subgrid Prescriptions for BH Accretion and Feedback}\label{sec:subgrid}

\begin{table*}[]
\centering
 \begin{tabular}{ c c c | c| c | c } 
{$R_B$ [$r_g$]} &  {n} & {$r_{\rm out}$ [$r_g$]}  &  {Time-averaged (\S~\ref{sec:subgrid1})} &{Distribution (\S~\ref{sec:subgrid2})} & Spin evolution (\S~\ref{sec:spin_evolution})\\[0.5ex]
 \hline\hline
{$\approx 400$} &  {4} & {$8^7\approx 2\times 10^6$} & \multirow{3}{*}{$
\overline{\dot{M}}(R_B,a_*)$~\eqref{eq:Mdot_powerlaw}\vspace{15em}} & \multirow{3}{*}{{$\ln\dot{M}\approxsim\mathcal{N}(\mu_{\ln\dot{M}}, \sigma_{\ln\dot{M}}^2$)
    }} &\multirow{5}{*}{spinup $\overline{s}(a_*)$~\eqref{eq:s_fit}} \\
{$\approx 2000$} &  {5} & {$8^9\approx 10^8$} & & & \\
{$\approx 2\times 10^4$} &  {6} & {$8^9\approx 10^8$} &   \multirow{3}{*}{$\overline{\dot{E}_{\rm fb}}(R_B, a_*)$~\eqref{eq:Efb_powerlaw}} & \multirow{3}{*}{{
    $\ln\eta\approxsim\mathcal{N}(\mu_{\ln\eta}, \sigma_{\ln\eta}^2$)}} &\\
{$\approx 2\times 10^5$} & {7} & {$8^{10}\approx 10^9$} & & &\\
{$\approx 2\times 10^6$} &  {8} & {$8^{11}\approx 10^{10}$} &  &  &\\
 \hline
 \end{tabular}
 \caption{Set-up of the multizone GRMHD simulations, and summary of the main results. The first three columns list the Bondi radius $R_B$, the number of zones $n$, and the outermost radius of the simulation $r_{\rm out}$. For each $R_B$, we perform simulations for 6 BH spin values, $a_*=0$, 0.1, 0.3, 0.5, 0.7, 0.9. The last three columns reference the sections and equations presenting the main results of this paper: two BH subgrid prescriptions (time-averaged quantities and their distributions) and the rate of evolution of BH spin.
 \label{tab:run_summary}}
\end{table*}

We have carried out BH accretion and feedback simulations with the multizone method for a range of Bondi radii $R_B/r_g$ and BH spin values $a_*$. The set-up of each simulation is described in \autoref{tab:run_summary}. There is a library of 30 simulations, which consists of 6 BH spin values\footnote{For the non-spinning case ($a_* = 0$), the run with the largest Bondi radius ($R_B = 2 \times 10^6\,r_g$) was evolved for a shorter duration of $500\,t_B$. Nevertheless, it reaches a final state consistent with our previous work \citep{Cho2024}.}, $a_* = 0$, 0.1, 0.3, 0.5, 0.7, 0.9, each with 5 choices of the Bondi radius, $R_B/r_g =400$, 2000, $2\times 10^4, ~2\times 10^5, ~2\times 10^6$. For $R_B/r_g = 2000$, we have additionally simulated an extremely high BH spin value of $a_*=0.97$. Snapshots of $a_*=0.9$ runs during the active jet episodes are shown in \autoref{fig:compare_snapshots}, where the relativistic jets remain collimated as they propagate well beyond several Bondi radii. From the simulations, we estimate the BH mass accretion rate $\overline{\dot{M}}/\dot{M}_B$ and the feedback power $\overline{\dot{E}}_{\rm fb}/\dot{M}_B c^2$, and derive subgrid prescriptions for each.

We present two versions of the BH accretion-feedback subgrid prescription from these GRMHD simulations. The first is a simple time-averaged version (Section~\ref{sec:subgrid1}), where we provide analytic formulae for $\overline{\dot{M}}(R_B,a_*)$ and  $\overline{\dot{E}_{\rm fb}}(R_B,a_*)$ as functions of the Bondi radius $R_B$ and the BH spin parameter $a_*$. The second prescription (Section~\ref{sec:subgrid2}) takes into account the fact that both accretion and feedback are highly time-variable; therefore, it provides information on the probability distributions of the accretion rate $\dot{M}$ and feedback efficiency $\eta$. Depending on the application, we anticipate that one or the other of these prescriptions will be more useful.

Note that our subgrid prescriptions depend on only two parameters: the dimensionless Bondi radius $R_B/r_g$ (or equivalently, the gas temperature $T_\infty$ beyond the Bondi radius, see \autoref{eq:RB}), and the BH spin parameter $a_*$. There is no dependence on any other gas properties outside the Bondi radius. As demonstrated in \citet{Cho2024,Cho2025} and in Appendix~\ref{sec:dependence_ic} here, the dynamical equilibrium state reached over a long runtime is independent of the initial gas configuration, gas rotation, or magnetic field strength. Thus, our BH subgrid models make no distinction between prograde and retrograde ICs, nor between strongly magnetized (MAD) and weakly magnetized (SANE) conditions.

\subsection{Prescription 1: Time-averaged accretion-feedback}\label{sec:subgrid1}

\begin{figure}[]
\gridline{
\fig{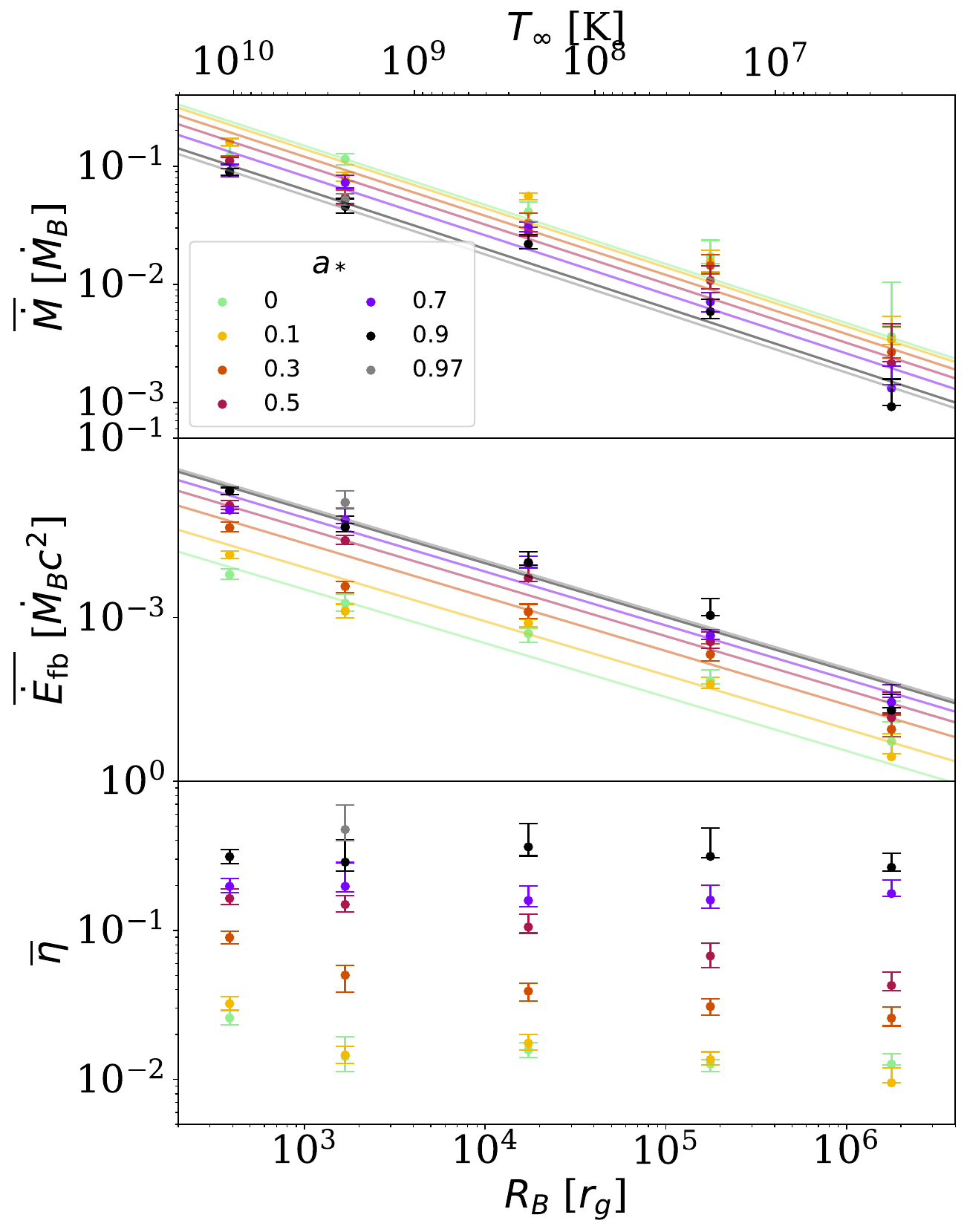}{0.48\textwidth}{}
            }
\caption{Time averaged accretion rate $\overline{\dot{M}}$ (top), feedback power $\overline{\dot{E}_{\rm fb}}$ (middle), and feedback efficiency $\overline{\eta}$ (bottom), for simulations with varying Bondi radii $R_B$ and BH spins $a_*$. The scale at the bottom shows the Bondi radius $R_B$ in units of the gravitational radius $r_g$, and the scale at the top shows the corresponding asymptotic temperature $T_\infty$. Colors represent different BH spins. The approximate formulae for $\overline{\dot{M}}$ (\autoref{eq:Mdot_powerlaw}) and $\overline{\dot{E}_{\rm fb}}$ (\autoref{eq:Efb_powerlaw}) are shown as straight lines in the top and middle panels.}\label{fig:compare_quantity_rB}
\end{figure}

\begin{figure*}[]
\gridline{
\fig{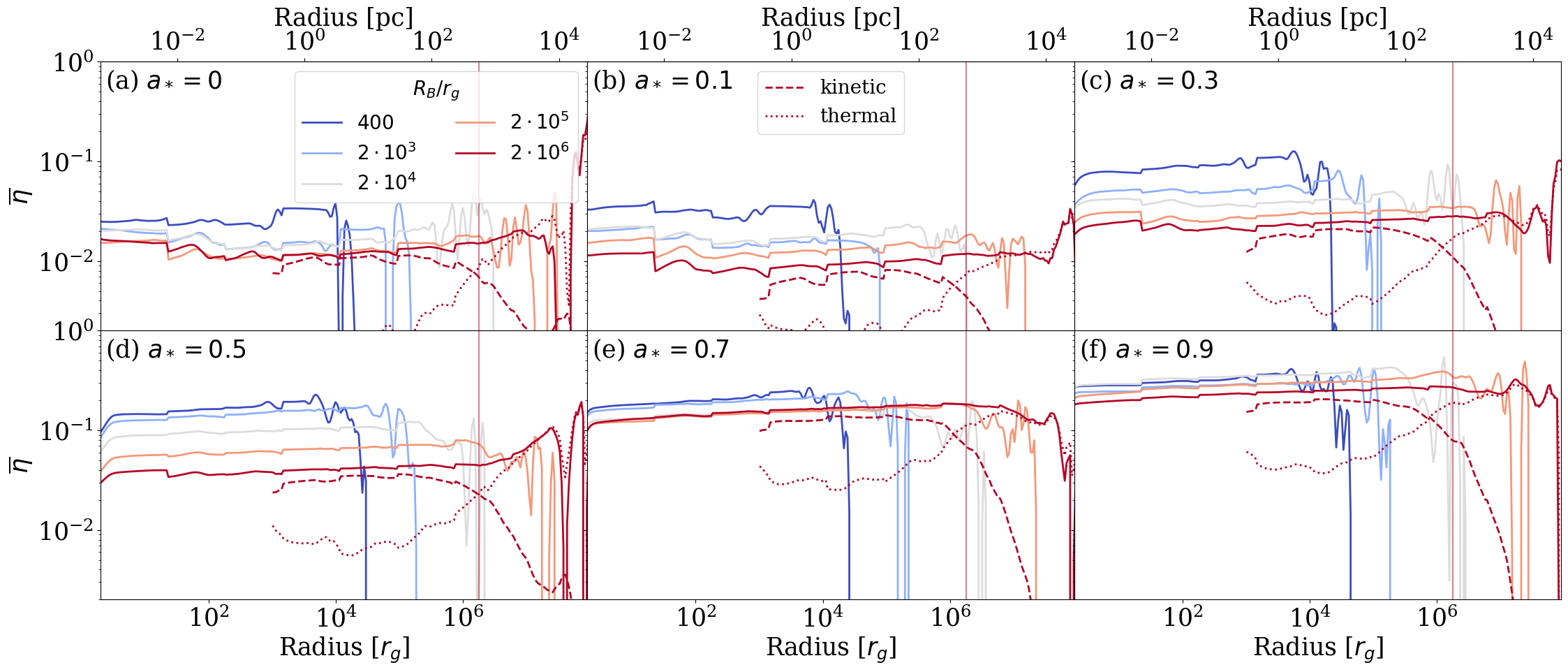}{1.\textwidth}{}
            }
\caption{Radial profiles of the time-averaged feedback efficiency $\overline{\eta(r)}$. Each panel corresponds to a single BH spin $a_*$ and colors represent different Bondi radii $R_B$. The scale at the bottom shows the radius in units of $r_g$, and the scale at the top shows the physical radius in pc for M87's BH mass. For the largest Bondi radius, $R_B = 2\times10^6\,r_g$, the kinetic $\overline{\eta_{\rm kinetic}(r)}$ (dashed) and thermal $\overline{\eta_{\rm thermal}(r)}$ (dotted) components are also overplotted, showing the conversion of kinetic to thermal energy as the jet outflow crosses $R_B$ (vertical lines).}\label{fig:eta_profiles}
\end{figure*}

The time-averaged accretion rate $\overline{\dot{M}}$, feedback power $\overline{\dot{E}_{\rm fb}}$, and feedback efficiency $\overline{\eta}$ for all runs are shown in \autoref{fig:compare_quantity_rB}, and their numerical values are listed in \autoref{tab:run_details}. The calculation of error bars is explained in \autoref{sec:lognormal}. In concordance with previous work, we find that the BH accretion rate scales with  the Bondi radius roughly as $\dot{M}/\dot{M}_B\appropto (R_B/r_g)^{-0.5}$. For a given Bondi radius $R_B$, there is a mild dependence of $\dot{M}$ on the BH spin, e.g., $\dot{M}$ is a factor of a few lower for $a_*=0.9$ compared to $a_*=0$, again consistent with previous works \citep[e.g.,][]{Guo2025, Cho2025v,Cho2025}. For $a_*=0.9$, we reproduce one of the main results of \citet{Cho2025}, namely, that the time-averaged feedback efficiency, $\overline{\eta}\approx 0.3$, is insensitive to the Bondi radius $R_B$. 

In the top panel of \autoref{fig:compare_quantity_rB}, there is evidence for some curvature in the variation of $\log(\overline{\dot{M}}/\dot{M}_B)$ vs $\log(R_B/r_g)$. However, for simplicity, we fit the measured values to straight lines. Furthermore, even though the best-fit lines have slightly different slopes, we assume the same slope for all spins, to obtain the following subgrid prescription\footnote{After fitting $\overline{\dot{M}}$ and $\overline{\dot{E}_{\rm fb}}$ assuming the same $R_B$ scaling for each BH spin, we fit their coefficients as a linear function of $a_*$ to obtain Equations~\ref{eq:Mdot_powerlaw} and \ref{eq:Efb_powerlaw}.} for the time-averaged mass accretion rate of the BH:
\begin{equation}\label{eq:Mdot_powerlaw}
    \frac{\overline{\dot{M}}}{\dot{M}_B} \approx (-3|a_*|+4.7) \left(\frac{R_B}{r_g}\right)^{-0.5}.
\end{equation}

Similarly, from the measurements shown in the middle panel of \autoref{fig:compare_quantity_rB}, we find that the time-averaged feedback power $\overline{\dot{E}_{\rm fb}}$ exhibits a similar power law scaling $\propto R_B^{-0.6}$ for all BH spins $a_*$. This gives the following approximate subgrid prescription for the feedback power,
\begin{equation}\label{eq:Efb_powerlaw}
    \frac{\overline{\dot{E}_{\rm fb}}}{\dot{M}_B c^2} \approx (0.98|a_*|+0.13)\left(\frac{R_B}{r_g}\right)^{-0.6}.
\end{equation}
These approximate subgrid prescriptions for $\overline{\dot{M}}$ and $\overline{\dot{E}_{\rm fb}}$ are shown as straight lines in the top and middle panels of \autoref{fig:compare_quantity_rB}.
Interestingly, our fit suggests that the feedback power depends linearly on the BH spin $a_*$, not quadratically as we would expect from force-free models of the jet \citep{Blandford1977,Tchekhovskoy2010,Tchekhovskoy2011}. While the reason is unclear, we note that the feedback power we are considering is the average of a highly time-variable quantity, whereas the analytical theory describes a perfectly time-steady phenomenon.

In the bottom panel of \autoref{fig:compare_quantity_rB}, the feedback efficiencies $\overline{\eta}$ are shown for all the runs. While rapidly spinning BHs have feedback efficiencies $\overline{\eta}$ independent of the Bondi radius, e.g., $\overline{\eta}\approx 0.3$ for $a_*=0.9$ and $\overline{\eta}\approx 0.2$ for $a_*=0.7$; slowly spinning BHs ($a_*\lesssim 0.5$) show a mild dependence on the Bondi radius, $\overline{\eta}\appropto (R_B/r_g)^{-0.2}$.\footnote{For the non-spinning case $a_*=0$, $\overline{\eta}$ seems to again have a very weak dependence on the Bondi radius $R_B$, with $\overline{\eta}\sim0.01$, consistent with our earlier work \citep{Cho2024}.}
It is possible that the weaker jets associated with slower spinning BHs have more difficulty competing with the large-scale inflowing matter and therefore perhaps there is a decline in feedback efficiency for larger $R_B$. Moreover, the slowest spinning case, $a_*=0.1$, exhibits nearly the same feedback efficiency as the  $a_*=0$ model, indicating that the jet is subdominant to the baseline level of feedback in the nonspinning case, which is associated with reconnection-driven convection \citep{Cho2023}.

Radial profiles of the time-averaged feedback efficiency $\overline{\eta(r)}$ are shown in Figure~\ref{fig:eta_profiles}, where each panel corresponds to a single BH spin $a_*$.\footnote{The jumps in feedback efficiency $\eta(r)$ across zones are reduced here compared to our previous work \citep[e.g., see Fig.~8 in][]{Cho2025}. This is due to our use of the improved inversion scheme of \citet{Kastaun2021} described in Appendix~\ref{sec:kastaun}, which avoids inversion failures.} For all runs, the feedback efficiency $\overline{\eta(r)}$ remains constant out to a few Bondi radii, indicating that each multizone simulation has achieved a global dynamical equilibrium. The previously noted difference in behavior between low and high BH spins is seen again in the $\overline{\eta(r)}$ profiles. The runs with slowly spinning BHs, $a_*=0.1$, 0.3, 0.5, have a wider spread in their $\overline{\eta}$ values with changing Bondi radius $R_B$, whereas for the runs with high BH spins, $a_*=0.7$, 0.9, the profiles are in close agreement between different $R_B$ for a given $a_*$.

At the Bondi radius $R_B$, kinetic and thermal components of the feedback efficiency $\overline{\eta}$ are nearly in equipartition, with $\overline{\eta_{\rm kinetic}}\approx\overline{\eta_{\rm thermal}}\approx\overline{\eta}/2$ for all runs. Inside $R_B$, the kinetic component dominates, while outside $R_B$, the thermal energy becomes dominant, as shown in \autoref{fig:eta_profiles}. 
This is not surprising, since the Bondi radius $R_B$ marks a transition point where the outflowing jet, which up to this point has propagated through a confining medium with a steeply declining pressure, suddenly encounters an extended layer of constant gas density and pressure. It is reasonable for the kinetic energy of the feedback to be converted to thermal energy at this point. However, it remains to be tested whether this behavior persists in realistic galactic environments beyond the Bondi radius.

\subsection{Prescription 2: Distribution of accretion-feedback}\label{sec:subgrid2}

\begin{figure*}[]
\gridline{
\fig{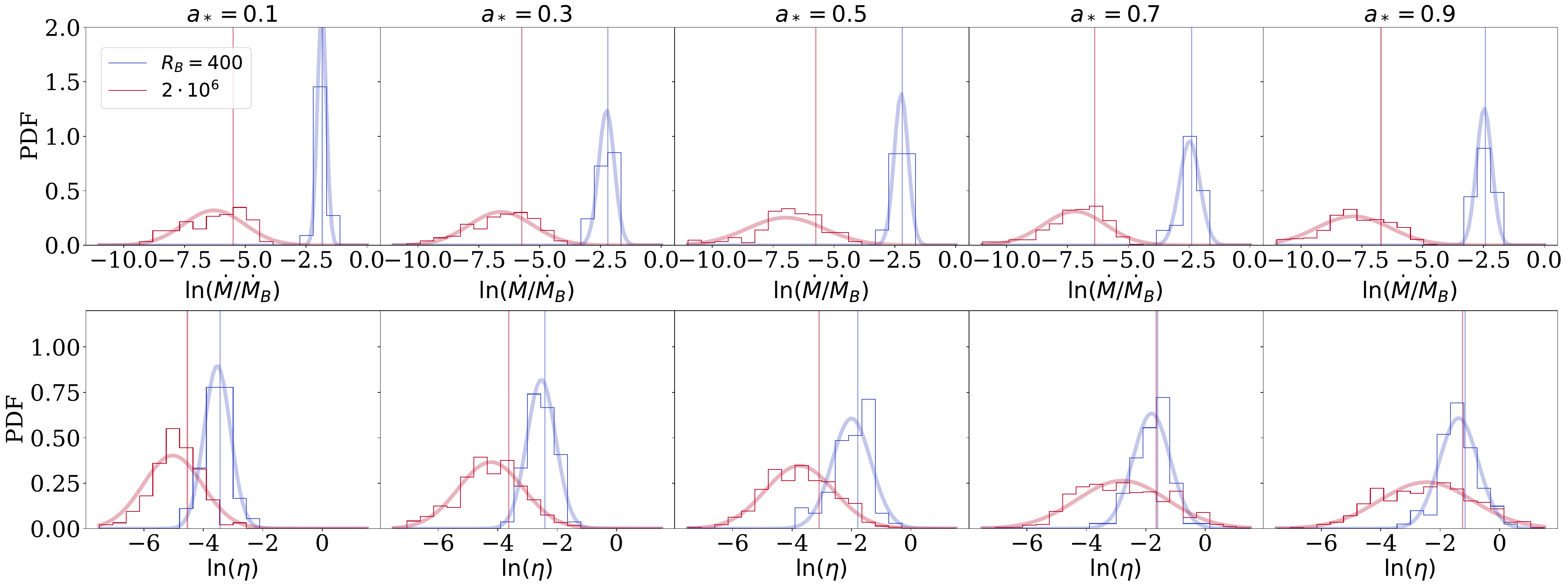}{1.\textwidth}{}
            }
\caption{Distributions of $\ln(\dot{M}/\dot{M}_B)$ (top row) and $\ln(\eta)$ (bottom row). Each column is for a given BH spin $a_*$ and the blue and red curves are for the smallest and largest Bondi radii, $R_B\approx 400\,r_g$ and $\approx 2\times 10^6\,r_g$, respectively. Lognormal curves corresponding to each run are overplotted for comparison. The means of the lognormal variable $X$, calculated using \autoref{eq:mean_lognormal} from log-space quantities $\mu_{\ln X}$ and $\sigma_{\ln X}$, are shown as vertical lines.}  \label{fig:eta_hist}
\end{figure*}

Here we present a more complete description of the statistical properties of $\dot{M}$ and $\eta$.
For spinning BHs ($a_*\neq 0$) with large Bondi radii $R_B$, the accretion rate $\dot{M}$ and feedback efficiency $\eta$ can fluctuate strongly by over 2 orders of magnitude \citep[][see also \citealt{Guo2025}]{Cho2025}. Therefore, the probability distributions of $\dot{M}$ and $\eta$  better represent the time-varying BH accretion and feedback, compared to the simple time-averages considered in Section~\ref{sec:subgrid1}.

The distributions of the logarithm of the accretion rate, $\dot{M}/\dot{M}_B$, and the feedback efficiency, $\eta$, are shown for a subsample of runs in \autoref{fig:eta_hist}. Both quantities have approximately lognormal distributions, with standard deviation $\sigma$ increasing with increasing Bondi radius $R_B$. The complete list of means and standard deviations of $\ln(\dot{M}/\dot{M}_B)$ and $\ln{(\eta)}$ is provided in \autoref{tab:run_details}.

Interestingly, for high BH spins, $a_*=0.7$, 0.9, the peak of the $\ln(\eta)$ distribution shifts substantially between $R_B\approx 400\,r_g$ and $R_B\approx 2\times 10^6\,r_g$ (see the two right panels in the bottom row). At first glance, this appears to contradict Section~\ref{sec:subgrid1}, where the time averaged efficiency $\overline{\eta}$ for these spin values was shown to be independent of the Bondi radius $R_B$. This apparent inconsistency can be understood through the properties of the lognormal distribution, as explained in Appendix~\ref{sec:lognormal}. If a statistical variable $X$ follows a lognormal distribution, the mean of $X$ is determined by a combination of the mean and standard deviation of $\ln X$ (\autoref{eq:mean_lognormal}). Therefore, the decreased mean $\mu_{\ln\eta}$ for large $R_B$ is compensated by an increased spread $\sigma_{\ln\eta}$, leading to similar average efficiencies $\overline{\eta}$. The estimated means calculated using \autoref{eq:mean_lognormal} are shown as vertical bars in \autoref{fig:eta_hist}. The values for $R_B=400\, r_g$ and $2\times10^6\, r_g$ indeed agree well for $a_*=0.7$ and 0.9, despite the two underlying distributions being very different. This is consistent with Section~\ref{sec:subgrid1}.

\section{BH Spin Evolution}\label{sec:spin_evolution}

\begin{figure}[]
\gridline{
\fig{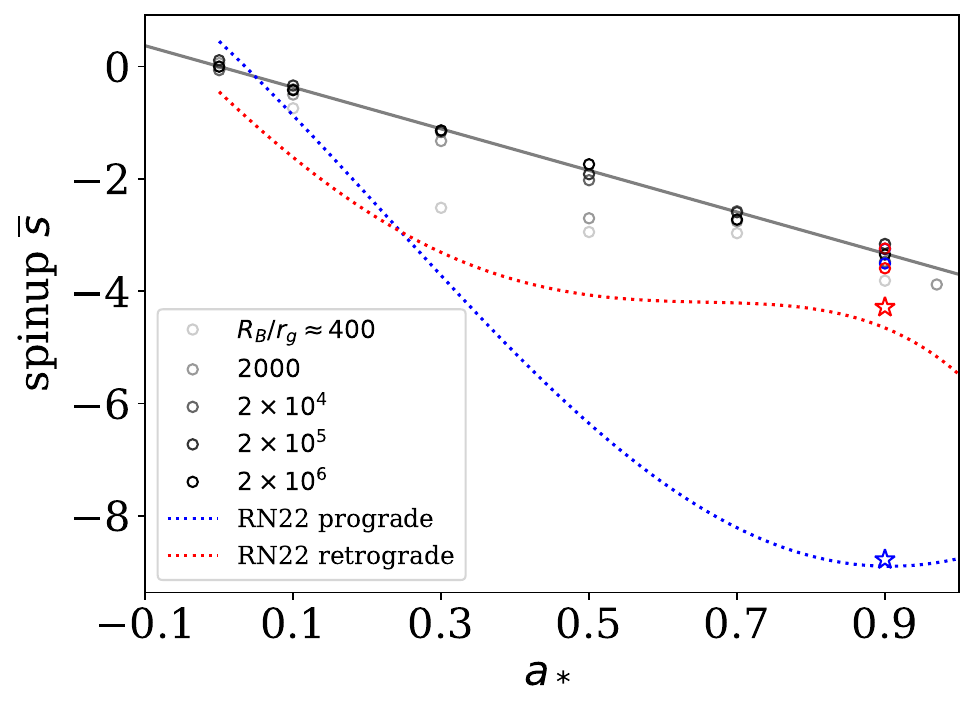}{0.48\textwidth}{}
            }
\caption{The spinup parameter $s$ (\autoref{eq:s_def}) as a function of BH spin $a_*$. Fiducial runs without initial gas angular momentum are shown in grayscale. Blue and red open circles represent runs with initially co- and counter-rotating gas, respectively, and agree well with the equivalent fiducial runs without rotation. \rn's small-scale GRMHD simulations are shown as dotted lines for prograde (blue) and retrograde (red). Multizone simulations limited to much shorter durations from \citet{Cho2025}, shown as stars, agree better with the \rn curves.
}\label{fig:compare_spinup}
\end{figure}

A feature of the subgrid prescriptions developed in this paper is that they depend on the BH spin parameter $a_*$. In order for cosmological simulations and SAMs to use these prescriptions, they will need to follow the spin evolution of SMBHs over cosmological timescales. For this, they need prescriptions for the rate of change of BH spin, $da_*/dt$, corresponding to each accretion scenario.

Our GRMHD simulations assume a fixed $a_*$ for each run, which is reasonable since the spindown timescale for the low Eddington ratio accretion we consider is much longer ($>1 {\rm Gyr}$, e.g., \citealt[][hereafter]{Narayan2022} \rn) than our simulation runtime $700\,t_B\approx50\,{\rm Myr}$ (assuming an M87-like system with a Bondi radius $R_B\approx 2\times 10^5\,r_g$). However, our simulations do provide information on $da_*/dt$ via the angular momentum inflow flux $\dot{J}$ defined in Equation \ref{eq:Jdot}.

The dimensionless spinup parameter $s$ is defined as \citep{Gammie2004,Shapiro2005}
\begin{equation}\label{eq:s_def}
    s\equiv {\frac{da_*}{dt}\frac{M}{\dot{M}}} = {j} - 2\, {e}\, a_*,
\end{equation}
where $j\equiv (\dot{J}/\dot{M})(5\,r_g)$ is the specific angular momentum inflow rate and $e\equiv (\dot{E}/\dot{M})(5\,r_g)$ is the specific energy inflow rate into the BH. In \autoref{fig:compare_spinup}, the time-averaged spinup parameters $\overline{s}$ corresponding to all runs are shown as gray dots, with light to bold shades signifying increasing Bondi radius $R_B$. All spinup parameters are negative, meaning that, in every case, the BH undergoes spindown.\footnote{Since there is no initial gas rotation, by symmetry, the same conclusion holds for negative spins: the BH spin slows down ($|a_*|$ decreases) in which case the spinup parameter is positive $s>0$.} Interestingly, runs with realistic Bondi radii $R_B = 2\times 10^4-2\times 10^6\,r_g$ give very similar spinup parameter values, so we can assume that $\overline{s}$ depends only on the BH spin $a_*$. Fitting a line through runs with $R_B= 2\times 10^4-2\times 10^6\,r_g$, we obtain the following simple formula for the spinup parameter,
\begin{equation}\label{eq:s_fit}
    \overline{s}(a_*) \approx -3.7 a_*,
\end{equation}
shown as the black line in \autoref{fig:compare_spinup}.

Additionally, we also show the runs with prograde and retrograde ICs discussed in Appendix~\ref{sec:dependence_ic}, with blue and red circles, respectively. The spinup parameters $s$ are similar for all ICs, regardless of the initial gas angular momentum. This again demonstrates the robustness of the final dynamical equilibrium, independent of the initial gas configuration \citep[][and Appendix~\ref{sec:dependence_ic}]{Cho2024,Cho2025} .

For completeness, we compare our results with conventional small-scale GRMHD simulations. \rn conducted a survey of a number of BH spins $a_*$, initialized with a small-scale torus of characteristic size $\sim 40\,r_g$ (radius of the pressure maximum). Their fitted spinup parameters (Equation 15 in \rn) are plotted in \autoref{fig:compare_spinup} as blue (prograde) and red (retrograde) dotted lines. Our multizone simulations do not agree with either of these curves.

The difference between our multizone runs and previous GRMHD simulations can be understood along the lines of the discussion in \citet{Cho2025v}. The runtime of most GRMHD simulations with torus ICs, including \rn, is less than or at best marginally equal to the time it takes the corresponding systems to reach their final state of strong variability.\footnote{\citet{Cho2025v} found that the onset time of strong variability increases with increasing initial rotation-to-magnetic energy ratio, $\R\equiv\rho r^2\Omega^2/b^2$. According to their prediction, the total runtime $10^5\,t_g$ of \rn is insufficient to reach the final state of strong variability.} On the other hand, when simulations are run for a much longer duration using the multizone method, the strongly variable state is always reached: the flow then alternates between prograde and retrograde states and the jet becomes intermittent. This explains the smaller values of $|\overline{s}|$ we find in our multizone simulations. Switching signs of the gas angular momentum leaves a negligible effect on the BH spin evolution. In addition, intermittency in the jet power makes extracting BH's spin energy less efficient. Together, these effects decrease the magnitude of the BH spindown parameter $|\overline{s}|$.

As a test, we compare our results with small-scale ($R_B\approx 400\,r_g$) multizone simulations from \citet{Cho2025} where the runtimes were reduced to $50\,t_B$ and magnetic fields were initially weak to reproduce typical GRMHD simulations (runs labeled as \texttt{mz+} and \texttt{mz-} in their work). The spinup parameters obtained from these runs are shown in \autoref{fig:compare_spinup} as blue and red stars for co-rotating and counter-rotating ICs, respectively. Indeed, when the runtime is much shortened, we recover spinup parameters similar to those of \rn's for prograde and retrograde. 
These results demonstrate that direct use of GRMHD simulations as subgrid prescriptions should be treated with caution, since typical GRMHD results probe much shorter timescales and may not capture the dynamical equilibrium reached over galactic timescales.

The BH spindown timescale can be calculated as
\begin{align}
    t_s \equiv \left|\frac{a_*}{\overline{da_*/dt}}\right| = \frac{a_*}{f_{\rm Edd}\frac{\dot{M}_{\rm Edd}}{M} |\overline{s}(a_*)|},
\end{align}
where the second equality uses the definition of $s$ in \autoref{eq:s_def}.
Here, $f_{\rm Edd}=\dot{M}/\dot{M}_{\rm Edd}$ is the Eddington ratio, and $M/\dot{M}_{\rm Edd} \equiv t_S$ is a BH mass independent constant, the Salpeter time $t_S = 4.5\times 10^7$\,yr.
Substituting \autoref{eq:s_fit} for $\overline{s}(a_*)$, we find that the spindown timescale depends only on the accretion rate as
\begin{equation}\label{eq:t_spindown}
    t_s = 12\left(\frac{10^{-3}}{f_{\rm Edd}}\right)\,{\rm Gyr},
\end{equation}
which is longer than the time scale of a few ${\rm Gyr}$ derived in previous GRMHD work \citep{Narayan2022, Ricarte2023}. 
For $f_{\rm Edd}\approx 10^{-3}$, the spindown timescale is already comparable with the Hubble time, and it is even longer for systems like M87 with $f_{\rm Edd} \lesssim 10^{-4}$.
This implies that BH spin evolution occurs mostly during periods of rapid accretion ($f_{\rm Edd}\sim 0.1-1$), or via BH mergers. When the BH enters the low-Eddington ratio phase $f_{\rm Edd}\lesssim 10^{-3}$, however, the BH spin is nearly fixed at a constant value until its next highly accreting phase.

\section{Conclusions \& Discussion}
\label{sec:conclusion}

Building on the multizone GRMHD framework developed in our earlier work, we have constructed a library of long-duration simulations that \emph{self-consistently} couples horizon-scale accretion physics to galactic scales across an unprecedented dynamic range. The suite spans BH spins $|a_\ast|=0$--0.9 and Bondi radii $R_B\simeq 4\times10^2$--$2\times10^6\,r_g$, enabling the systematic derivation of first-principles, spin-dependent subgrid prescriptions appropriate to hot, radiatively inefficient accretion flows at low Eddington ratios ($f_{\rm Edd}\lesssim10^{-3}$).

From this parameter study, we provide two complementary subgrid models for cosmological simulations and SAMs:
\begin{itemize}
\item {Time-averaged prescriptions:} We supply compact analytic fits for the suppressed mean accretion rate and the mean feedback power, normalized to the classic Bondi rate $\dot M_B$ (cf.\ \autoref{eq:MdotB}). In our simulations, both $\overline{\dot M}/\dot M_B$ and $\overline{\dot E}_{\rm fb}/(\dot M_B c^2)$ follow simple power-law scalings with $R_B$, with coefficients that vary linearly with spin (Equations~\ref{eq:Mdot_powerlaw} and \ref{eq:Efb_powerlaw}):
\begin{align*}
\frac{\overline{\dot M}}{\dot M_B} &\approx \left(-3|a_\ast|+4.7\right)\left(\frac{R_B}{r_g}\right)^{-0.5}, \\
\frac{\overline{\dot E}_{\rm fb}}{\dot M_B c^2} &\approx \left(0.98|a_\ast|+0.13\right)\left(\frac{R_B}{r_g}\right)^{-0.6}.
\end{align*}
A key qualitative outcome is that higher-spin BHs deliver larger feedback power at fixed $R_B$, while exhibiting modestly {lower} accretion rates, consistent with a stronger jet/outflow regulating inflow on large scales.

\item {Stochastic (distribution-level) prescriptions:} As accretion and feedback are intrinsically and strongly time-variable---especially for larger $R_B$ and nonzero spin---we additionally characterize the \emph{full} probability distributions of $\dot M$ and $\eta$. Both are well described by lognormal statistics, with the width (scatter in $\ln \dot M$ and $\ln\eta$) increasing with $R_B$. We tabulate the corresponding means and variances for all models in \autoref{tab:run_details}, enabling applications that require bursty coupling rather than relying solely on time-averaged properties.
\end{itemize}

Since our accretion-feedback prescription is explicitly spin-dependent, we also quantify BH spin evolution in the low-$f_{\rm Edd}$ regime. Using the angular-momentum fluxes measured in the simulations, we find that the spinup parameter is well approximated by a simple linear function of $a_*$ (\autoref{eq:s_fit}),
\begin{equation*}
s(a_\ast)\simeq -3.7\,a_\ast,
\end{equation*}
implying spindown in this hot-mode state. The associated spindown timescale is extremely long (\autoref{eq:t_spindown}),
\begin{equation*}
t_s \simeq 12\left(\frac{10^{-3}}{f_{\rm Edd}}\right)\,{\rm Gyr},
\end{equation*}
so for $f_{\rm Edd}\lesssim10^{-3}$ the spin is effectively ``frozen'' over cosmological times. Thus, spin changes are expected to occur primarily during high-accretion episodes and/or mergers, not during quiescent ADAF phases.

The subgrid prescription developed in Section~\ref{sec:subgrid1} is designed for direct use in cosmological simulations and SAMs, provided the BH is in the low-Eddington, hot-accretion regime ($f_{\rm Edd}\lesssim 10^{-3}$). At higher accretion rates, where the flow is expected to transition to a radiatively efficient state, a different accretion/feedback model should be adopted. We envision the implementation of these schemes into simulations \& SAMs proceeding as follows:

\begin{enumerate}
    \item {Estimate the Bondi scale:} Determine the Bondi radius in gravitational units, $R_B/r_g$, by measuring the average gas temperature $T_\infty$ in the BH’s vicinity as resolved in cosmological models (which typically exceed the Bondi radius scales) via \autoref{eq:RB}. In practice, this step highlights a common limitation of large-volume runs: the Bondi radius is often only marginally resolved, and the inferred $T_\infty$ (hence $R_B$) can be sensitive to the adopted measurement aperture and subgrid ISM treatment.

    \item {Compute the Bondi inflow rate:} Measure the gas density $\rho$ at the Bondi scale (or estimate it from the local density if unresolved) and compute the Bondi accretion rate $\dot{M}_B$ using \autoref{eq:MdotB}.

    \item {Apply the suppression of accretion:} Use the calibrated scaling in \autoref{eq:Mdot_powerlaw} to obtain the \emph{suppressed} accretion rate $\dot{M}$ relative to $\dot{M}_B$, and evolve the BH mass by $\dot{M}$.

    \item {Inject feedback power:} Compute the feedback power $\dot{E}_{\rm fb}$ from \autoref{eq:Efb_powerlaw} and couple it to the surrounding gas. If possible, a collimated (bipolar) injection geometry is the most physically motivated choice for hot-mode feedback (see \autoref{fig:compare_snapshots}).

    \item {Evolve the BH spin:} Update the BH spin $a_\ast$ using the fitted spin-evolution relation in \autoref{eq:s_fit}.
 \end{enumerate}

Finally, our results, together with \citet{Cho2025}, underscore a central cautionary lesson: short-duration, small-domain GRMHD simulations can mis-estimate long-term, galaxy-scale behavior when directly repurposed as subgrid models. The multizone approach reaches a global dynamical equilibrium over scales extending beyond $R_B$, and captures the strong variability of the equilibrated state (including jet intermittency and an effective loss of initial rotational memory) that materially alters the inferred mean energetics and spin evolution compared to conventional small-scale GRMHD simulations. This is precisely what makes the present prescriptions robust and directly portable: the feedback power measured near the Bondi radius $R_B$ is obtained in a self-consistent, equilibrated inflow--outflow system and, therefore, can be transferred to larger-scale cosmological calculations without uncontrolled extrapolation.

We note some caveats of this work. Here, we considered BH spins in the range $|a_*|=0-0.9$, so the formulae for $\dot{M}(R_B,a_*)$, $\dot{E}_{\rm fb}(R_B,a_*)$, and $s(a_*)$ (Equations~\ref{eq:Mdot_powerlaw}, \ref{eq:Efb_powerlaw}, \ref{eq:s_fit}) are not validated for near-maximal BH spins.
In addition, because our GRMHD framework neglects radiative effects, the resulting BH subgrid prescriptions are applicable only to low-$\dot{M}$ ADAF states, as radiative cooling may already influence accretion flows starting at $\dot{M} \gtrsim 10^{-5}\dot{M}_{\rm Edd}$ \citep{Singh2025}. For discussions of the accretion rate $\dot{M}$ threshold below which ADAF states are expected, see, e.g., \citet{Yuan2014,Cho2022}. 
Finally, we adopt primarily vertical magnetic field geometries. Although large-scale poloidal fields are expected to arise naturally in realistic environments \citep[e.g.,][]{Liska2020}, alternative magnetic configurations (e.g., toroidal or randomized) deserve further study.
The caveats of the multizone method itself are described extensively in \citet{Cho2024,Cho2025} and are not repeated here. 

By bridging horizon-scale physics to galactic scales via the multizone method, this work establishes a physically grounded foundation for spin-dependent BH subgrid models in next-generation cosmological simulations and SAMs.
These prescriptions for low accretion modes will be particularly useful for models that explicitly track BH spin evolution (e.g., NEWHORIZON \citealt{Dubois2021,Beckmann2025}, AREPO \citealt{Talbot2021}, COLIBRE \citealt{Schaye2025, Husko2025}).

\section*{Acknowledgements}
H.C., B.P., R.N., K.S., A.R., and P.N. were partially supported by the Black Hole Initiative at Harvard University, which is funded in part by the Gordon and Betty Moore Foundation (Grant \#13526). It was also made possible through the support of a grant from the John Templeton Foundation (Grant \#63445). The opinions expressed in this publication are those of the author(s) and do not necessarily reflect the views of these Foundations.
This publication is funded in part by the Gordon and Betty Moore Foundation, Grant GBMF-12987.
A.P.V. gratefully acknowledges support from the National Science Foundation Astronomy \& Astrophysics Postdoctoral Fellowship under Award Number 2502826.
This work used Delta at the University of Illinois at Urbana Champaign through allocation PHY250303 from the Advanced Cyberinfrastructure Coordination Ecosystem: Services \& Support (ACCESS) program, which is supported by U.S. National Science Foundation grants \#2138259, \#2138286, \#2138307, \#2137603, and \#2138296.

\clearpage
\newpage

\appendix

\section{Overview of the Multizone Method and Additional Numerical Details}\label{sec:method_detail}
\subsection{Multizone Method}
In the multizone method, the full simulation domain, extending from $r=r_g$ to $r_{\rm out}$, is divided into different subregions, or zones, and evolved in a manner similar to a ``V-cycle''.
Each zone-$i$ has a different inner radius, $r_{i,\rm in}=8^i\,r_g$, and a common outer radius, $r_{i,\rm out}=r_{\rm out}$, where $r_{\rm out}$ is chosen to be much larger than the Bondi radius $R_B$ (see Table~\ref{tab:run_summary}). For $n$ zones, the method begins by evolving the outermost zone-$(n-1)$. Then, the active zone is switched sequentially to zone-$(n-2)$, zone-$(n-3)$,  \ldots, down to zone-$0$, which incrementally moves the inner boundary inwards down to $r=r_g$. This constitutes the inward moving half of the V-cycle, which tracks accretion. The inner boundary is then moved back out from zone-$0$ to zone-$(n-1)$, completing one full V-cycle. The second half of the V-cycle captures the outflowing part of the communication or feedback. For a given simulation, V-cycles are repeated over a few tens to hundreds of times to ensure enough communication between small and large scales (see the third column of \autoref{tab:run_details} for the total number of V-cycles for each run).
Following \citet{Cho2025}, each zone is evolved for a duration of $8000\Delta t$ before switching to the next zone, where $\Delta t$ is the on-the-fly calculated timestep set by the Courant condition for that zone. For zone-$0$, which covers the whole simulation domain, we run 10 times longer ($80000\Delta t$) to allow for extra relaxation. Tests of different choices of the zone boundaries and runtime per zone are described in Appendix F of \citet{Cho2024}.
 
Besides the usual simulation boundaries at $r_g$ and $r_{\rm out}$, the multizone scheme introduces additional internal radial boundaries at $r_{i>0,\rm in}$. These internal boundaries are assigned Dirichlet boundary conditions such that $\rho$, $u$, $u^\mu$, and $B^i$ are held constant over the duration of each zone. Importantly, for the magnetic fields, we apply an additional prescription, labeled \bfluxc{} in \citet{Cho2024,Cho2025}, to the electric fields on the internal radial boundary face. This prescription ensures the divergence-free condition, $\nabla\cdot \vec{B} =0$, and avoids destructive magnetic tension between active and inactive regions. For implementation details and tests of the \bfluxc{} prescription, see Appendix E of \citet{Cho2024} and Section 3 of \citet{Cho2025}.
At the innermost radius, $r=r_{0,\rm in}=r_g$, which lies inside the BH horizon, the gas is allowed to flow freely towards the singularity. Similarly, gas can freely leave the simulation domain at the outermost radius, $r=r_{\rm out}$. In the $\theta$ and $\varphi$ directions, we use reflecting and periodic boundary conditions, respectively.

The key strength of the multizone method is that it significantly lowers the computational cost of these large dynamic range simulations. This is achieved by periodically freezing and unfreezing the inner scales inside $r_{i,\rm in}$, thereby permitting a larger timestep suitable for the currently active radial range. The multizone method's careful design of progressively moving the interior boundary $r_{i,\rm in}$ in and out allows each scale to respond to any information transferred from both inner and outer scales. Thus, a global dynamical equilibrium is quickly reached with a reasonable computational cost.

\subsection{Other Numerical Set-up}
The numerical set-up in the present work is identical to that in \citet{Cho2025}, except for the inversion scheme. In \citet{Cho2025}, the 1D$_W$ inversion scheme \citep{Noble2006, Mignone2007} was used, where occasional inversion failures at the internal boundaries resulted in jumps in the radial profile for the efficiency $\eta(r)$. Here, we adopt the more robust inversion scheme of \cite{Kastaun2021} which avoids such failures. We made some modifications to their original \citet{Kastaun2021} scheme to avoid problematic high velocity cells during the inversion (details are given in Appendix~\ref{sec:kastaun}). We adopt face-centered magnetic fields and the same set of floors as in \citet{Cho2025}.

We adopt an exponential Kerr-Schild (EKS) coordinate system where the grid is evenly spaced in $\log{r}$, $\theta$, $\varphi$. This naturally focuses the spatial resolution closer to the BH. To avoid unnecessarily short timesteps from small $\varphi$ edges of cells near the polar axes ($\theta=0,\pi$), we use an internal static mesh refinement (ISMR) scheme that de-refines cells in the $\varphi$ direction. The ISMR implementation details are presented in \citet{Cho2025}. For the resolution of $N_r\times N_\theta \times N_\varphi = 32\log_{8}(r_{\rm out}/r_g) \times 64\times 64$, where $r_{\rm out}$ is the outermost radius of the simulation, ISMR is used on 4 layers of the $\theta$ grid near each pole.

Time-averaged quantities are calculated over the latter half of the total runtime (i.e., $t = 350-700\,t_B$). The time-averages are calculated for each zone individually, using the final snapshot of each zone before switching zones, in order to minimize the imprint of internal boundaries. Then, we combine zone-$i$'s time-averages at radii $8^{0.5}\,r_{i,\rm in} - 8^{1.5}\,r_{i,\rm in}$, except for zone-$0$ and zone-$(n-1)$ where the range extends to the innermost radius $r_{0,\rm in}$ and outermost radius $r_{\rm out}$ of the simulation, respectively.

\subsection{Momentum Conserving Inversion Scheme}\label{sec:kastaun}
The technical improvement of this work relative to \citet{Cho2025} is the adoption of the \citet{Kastaun2021} inversion scheme that prevents discontinuities in the outflow power across zones. Here we describe our slightly modified implementation of the scheme.
In \citet{Kastaun2021}, the primitives $\mathbb{P}=\{\rho,u,v^i,B^i\}$ are inverted from conserved quantities in the Eulerian frame: density $-\rho u^\mu n_\mu = \rho \Gamma$, net energy density $n_\alpha n_\beta T^{\alpha\beta}+\rho u^\alpha n_\alpha$, three-momentum density $-\gamma^\mu_i n^\alpha T_{\alpha\mu}=\alpha T^t_i$, and magnetic fields. For a given set of conserved variables $\{\rho\Gamma, n_\alpha n_\beta T^{\alpha\beta}+\rho u^\alpha n_\alpha,\alpha T^t_i,B^i\}$, the scheme finds a unique solution of $\mathbb{P}$. While the resulting solution is reasonable in most cases, the inversion occasionally produces extremely low internal energy $u$ and extremely fast velocities $v^i$, in the strongly magnetized regions $\beta\ll 1$.
In such problematic cases where the \citet{Kastaun2021} scheme inverted velocity exceeds the desired maximum Lorentz factor $\Gamma_{\rm max}$, we instead apply the following inversion that conserves only the momentum density $T^t_i$ and magnetic fields $B^i$, sacrificing conservation of density and net energy density -- that is, inserting new density and internal energy in the coordinate frame. The new scheme is described below.

First, we separate the momentum density $S_i=\alpha T^{t}_i$ into parallel and perpendicular components with respect to magnetic fields.
\begin{align}
    S_\parallel^i = \rho h \Gamma^2 v_\parallel^i, \quad S_\perp^i = (\rho h \Gamma^2+B^2)v_\perp^i,
\end{align}
where $h=1 + \gamma_{\rm ad} u/\rho$ is the specific enthalpy, $v_\parallel$ and $v_\perp$ are parallel and perpendicular velocities to magnetic fields respectively.
Then, when expressing the magnitude of velocity in terms of the momentum density,
\begin{align}
    v^2 = \frac{S_\parallel^2}{(\rho h\Gamma^2)^2} + \frac{S_\perp^2}{(\rho h \Gamma^2 + B^2)^2},
\end{align}
the above is equal to $1- 1/\Gamma^2$. Then we solve for the function
\begin{align}
    f(\Gamma) = \frac{S_\parallel^2}{(\rho h\Gamma^2)^2} + \frac{S_\perp^2}{(\rho h \Gamma^2 + B^2)^2} + \frac{1}{\Gamma^2}-1=0.
\end{align}
The function $f(\Gamma)$ has a nice property that it monotonically decreases $f'(\Gamma)<0$ and $f(1)>0$. Therefore, in order for $f(\Gamma)$ to have a solution in the range $\Gamma\in[1,\Gamma_{\rm max}]$, we need to ensure that $f(\Gamma_{\rm max}) < 0$. For a given $S_\parallel$, $S_\perp$, $\vec{B}$, the condition $f(\Gamma_{\rm max})<0$ constrains the value of $\rho h$.
If we assert that
\begin{align}
    f(\Gamma_{\rm max}) < \frac{S_\parallel^2}{(\rho h)^2\Gamma_{\rm max}^4} + \frac{S_\perp^2}{(\rho h)^2 \Gamma_{\rm max}^4} + \frac{1}{\Gamma_{\rm max}^2}-1 <0,
\end{align}
$\rho h$ should be greater than
\begin{align}
    \rho h > \frac{S}{\Gamma_{\rm max}^2 \sqrt{1-1/\Gamma_{\rm max}^2}}=(\rho h)_{\min}.
\end{align}

The implementation order of the modified \citet{Kastaun2021} scheme is as follows. After \citet{Kastaun2021} inversion, if $\rho h$ is smaller than $(\rho h)_{\rm min}$, $\rho$ and $u$ are increased proportionally to conserve the temperature
\begin{equation}
    \rho \to \rho \frac{(\rho h)_{\rm min}}{\rho h}, \quad u \to u \frac{(\rho h)_{\rm min}}{\rho h}.
\end{equation}
With the newly adjusted $\rho h$, the function has a value $f(\Gamma_{\rm max})<0$, so there is a unique solution $\Gamma$ to the equation $f(\Gamma)=0$.
Finally, we recover velocities using the solution. 
\begin{align}
    v^i &= v_\parallel^i + v_\perp^i\\
    &=\frac{S_\parallel^i}{\rho h \Gamma^2} + \frac{S_\perp^i}{\rho h \Gamma^2+B^2}\\
    &=\frac{S_\parallel^i}{\rho h \Gamma^2} + \frac{S^i-S_\parallel^i}{\rho h \Gamma^2+B^2},
\end{align}
where $S_\parallel^i=(\vec{B}\cdot\vec{S})B^i/B^2$.

\section{Dependence on Initial Conditions}\label{sec:dependence_ic}
\begin{figure}[]
\gridline{
\fig{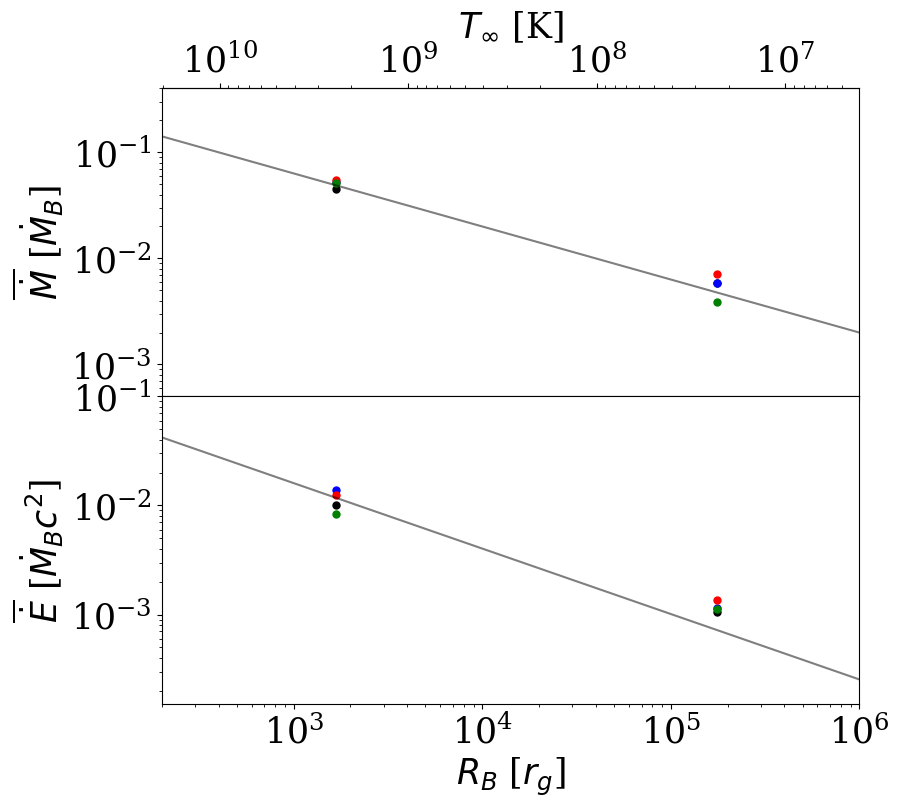}{0.48\textwidth}{}
            }
\caption{Time-averaged accretion rate $\overline{\dot{M}}$ and feedback power $\overline{\dot{E}_{\rm fb}}$ for the $a_*=0.9$ runs with different ICs: fiducial non-rotating (black), co-rotating (blue), counter-rotating (red), and weakly magnetized (green). For reference, the fitted relations $\overline{\dot{M}}(R_B,a_*=0.9)$ and $\overline{\dot{E}_{\rm fb}}(R_B,a_*=0.9)$ (Equations~\ref{eq:Mdot_powerlaw} and \ref{eq:Efb_powerlaw}) are shown as black lines.}\label{fig:compareic}
\end{figure}

Here we demonstrate the robustness of our subgrid prescriptions to different ICs, specifically gas rotation and magnetic field strength.
Traditionally, GRMHD simulations distinguish prograde ($a_*>0$) and retrograde ($a_*<0$) BH spins and adopt a highly idealized \citet{Fishbone1976}-type torus IC with dynamically subdominant magnetic fields. More recently, however, GRMHD studies have increasingly explored more general classes of ICs and found qualitatively different accretion states \citep[e.g.,][]{Ressler2020b,Ressler2021,Lalakos2022,Lalakos2025,Cho2023,Cho2024,Cho2025,Cho2025v,Kaaz2023,Kaaz2025,Kwan2023,Galishnikova2025,Guo2025,Chan2025,Kim2025}. 

Among these studies, only a few have reported flips in rotation direction in the presence of strong magnetic fields \citep{Cho2024,Cho2025,Cho2025v,Galishnikova2025}, challenging the traditional distinction between prograde and retrograde states and instead suggesting that realistic accretion flows may alternate between the two over time. Using the multizone method, \citet{Cho2025} has successfully demonstrated that when evolving \citet{Fishbone1976}-like IC (denoted {\bf T+} and {\bf T-} ICs in their paper) for extended durations with a Bondi radius of $R_B\approx 400\,r_g$, the coherent sense of gas rotation is lost as magnetic fields accumulate and become dynamically important. Therefore, the resulting time-averaged feedback efficiency ($\eta\sim0.3$) differs from that of typical GRMHD simulations, which yield $\eta\sim 1$ for prograde and $\eta\sim0.1$ for retrograde.

Here we further demonstrate the independence of the final state on different ICs for even larger Bondi radii of $R_B \approx 2000\,r_g$ and $\approx 2\times 10^5\,r_g$. We repeat $a_*=0.9$ simulations with 3 different types of ICs: prograde, retrograde, and weakly magnetized. Each new ICs differs from the fiducial Bondi-like IC ({\bf B} in \citealt{Cho2025}) by a single feature. In the prograde and retrograde ICs, the gas co-rotates and counter-rotates with the BH, respectively, with azimuthal velocity $|u^\varphi| = 0.5 r^{-3/2}$, respectively. For the weakly magnetized IC, the initial plasma-$\beta$ is increased from $1$ to $100$. The three additional large Bondi radius $R_B \approx 2\times10^5\,r_g$ runs are evolved for $400\,t_B$,shorter than the fiducial runtime but still substantial. For reference, the rotational-to-magnetic energy ratio, $\R\equiv\rho r^2\Omega^2/b^2$ \citep{Cho2025v}, measured at $R_B$ is $\R\sim 0.2$ for prograde/retrograde ICs and $\R=0$ for the fiducial (Bondi-like) and weakly magnetized ICs.

The final time-averages of the accretion rate $\overline{\dot{M}}$ and feedback power $\overline{\dot{E}_{\rm fb}}$ are shown in \autoref{fig:compareic}. The runs with fiducial Bondi-like ICs are shown as black dots, while the prograde, retrograde, and weakly magnetized IC runs are shown in blue, red, and, green, respectively. The three new types of ICs are in good agreement with the fiducial run in both $\overline{\dot{M}}$ and $\overline{\dot{E}_{\rm fb}}$. This confirms our previous finding that the final dynamical equilibrium is insensitive to the initial gas dynamics and magnetic field strengths.

\section{Detailed Properties of Individual Simulations}\label{sec:details}

\begin{table*}[]
\centering
 \begin{tabular}{c c| c  |  c c | c c c c} 
\multirow{2}{*}{$R_B$ [$r_g$]} & \multirow{2}{*}{BH spin $a_*$} & \multirow{2}{*}{$N_{\rm V-cycles}$} &  \multicolumn{2}{c|}{Prescription 1 (\S~\ref{sec:subgrid1})} &\multicolumn{4}{c}{Prescription 2 (\S~\ref{sec:subgrid2})}\\\cline{4-5}\cline{6-9}
& & &$\overline{\dot{M}}$ [$\dot{M}_B$] & $\overline{\eta}$ & $\mu_{\ln\dot{M}}$ & $\sigma_{\ln{\dot{M}}}$ & $\mu_{\ln\eta}$ & $\sigma_{\ln{\eta}}$ \Tstrut\\[0.5ex] 
 \hline\hline
\multirow{6}{*}{$\approx 400$} &  0 & 33  & 0.12 &  0.026 &  -2.2 & 0.41 & -3.8  & 0.47 \\
 &  0.1 & 38 & 0.16  & 0.032 & -1.9  & 0.18 & -3.5  & 0.45 \\
 &  0.3 & 59 & 0.11  & 0.089 & -2.3 & 0.32& -2.5 & 0.49 \\
&  0.5 & 76 & 0.11 & 0.16 & -2.2 & 0.29 & -2.0 & 0.66 \\
 &  0.7 & 78 & 0.091 & 0.20 & -2.5  & 0.42 & -1.8 & 0.63\\
 &  0.9 & 86 & 0.090 & 0.31 & -2.5 & 0.32 & -1.4 & 0.66 \\
 \hline
\multirow{7}{*}{$\approx 2000$} &  0 & 15 & 0.11 & 0.014& -2.2 &0.21 & -4.4 & 0.67 \\
 &  0.1 & 15 & 0.081  & 0.015 & -2.5 & 0.17 & -4.3 & 0.36 \\
 &  0.3 & 24 & 0.055 & 0.050 & -3.0 & 0.33 & -3.3 & 0.69 \\
 &  0.5 & 50& 0.053 &  0.15 & -3.0 & 0.35 & -2.1 & 0.62 \\
&  0.7 & 44 & 0.072 &  0.20 & -2.7 & 0.45 & -2.0 & 0.98 \\
 &  0.9 & 48 & 0.045 &  0.29 & -3.2 & 0.50 & -1.7 & 1.1 \\
 &  0.97 & 50 & 0.052 &  0.47 & -3.0 & 0.35 & -1.3 & 1.2 \\
 \hline
\multirow{6}{*}{$\approx 2\times 10^4$} &  0 & 45  &  0.041 & 0.016 & -3.4 & 0.60 & -4.3 & 0.52 \\
 &  0.1 & 46 & 0.056 &  0.018 & -3.0 & 0.35 & -4.2 & 0.59 \\
 &  0.3 & 65 & 0.033 &  0.039 & -3.6 & 0.64 & -3.5 & 0.70\\
 &  0.5 & 105 & 0.028 &  0.11 & -3.7 & 0.48 & -2.7 & 0.95 \\
 &  0.7 &138 & 0.030 &  0.16 & -3.6 & 0.53 & -2.4 & 1.1 \\
 &  0.9 & 124 & 0.022 &  0.36 & -4.0 & 0.73 & -2.0 & 1.5 \\
 \hline
\multirow{6}{*}{$\approx 2\times 10^5$} &  0 & 105 & 0.017 & 0.013 & -4.5 & -0.99  & -4.6 &  0.64 \\
 &  0.1 & 109 & 0.016 & 0.014 & -4.5 & 0.78 & -4.5 & 0.70\\
 &  0.3 & 156 & 0.014 &  0.031 & -4.6 & 0.83 & -3.8 & 0.84 \\
 &  0.5 & 181 & 0.011 & 0.067 & -4.9 & 0.91 & -3.4 & 1.1 \\
 &  0.7 & 258 & 0.0071 & 0.16 & -5.4 & 0.93 & -2.6 & 1.3 \\
 &  0.9 & 262 & 0.0059 &0.31 & -5.6 & 1.0  & -2.1 & 1.5 \\
 \hline
\multirow{6}{*}{$\approx 2\times 10^6$} & 0 & 164 & 0.0036 & 0.013 & -6.3 & 1.6 & -4.6 & 0.80 \\
 & 0.1 & 210 & 0.0034 & 0.0095 & -6.3 & 1.3 & -5.0 & 0.99 \\
 &  0.3 & 224 & 0.0027 & 0.026 & -6.6 & 1.3 & -4.2 & 1.1 \\
 &  0.5 & 322 & 0.0021 & 0.043 & -7.0 & 1.6 & -3.8 & 1.2 \\
 &  0.7 & 519 & 0.0013 & 0.18 & -7.2 & 1.3 & -2.8 & 1.5 \\
 &  0.9 & 523 & 0.00092 & 0.26 & -7.8 & 1.4 & -2.5 & 1.6 \\
 \hline
 \end{tabular}
 \caption{Detailed information on BH accretion and feedback for all simulations. The first three columns list the Bondi radius $R_B$, BH spin $a_*$, and the total number of V-cycles. Columns 4 and 5 give time-averaged accretion rate $\overline{\dot{M}}$ and the feedback efficiency $\overline{\eta}$ in Section~\ref{sec:subgrid1}. The last four columns show the mean $\mu$ and standard deviation $\sigma$ of $\ln{(\dot{M}/\dot{M}_B)}$ and $\ln{\eta}$ relevant to Section~\ref{sec:subgrid2}.
 \label{tab:run_details}}
\end{table*}

\begin{figure*}[]
\gridline{
\fig{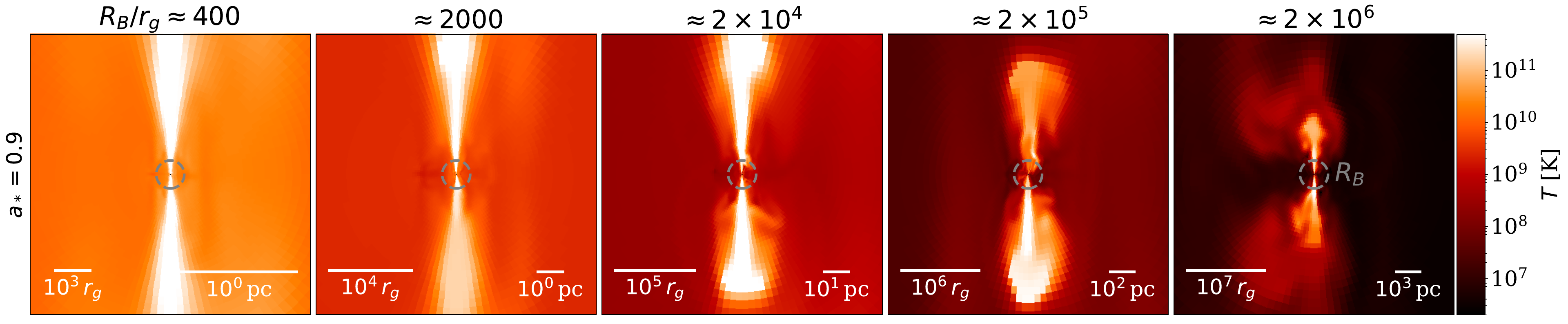}{1.\textwidth}{}
            }
\caption{Temperature snapshots of the runs with BH spin $a_*=0.9$. From left to right, the Bondi radius $R_B$ (shown as gray dotted circles) increases, or equivalently, the asymptotic gas temperature $T_\infty$ decreases. The horizontal bars in the left bottom corner of each panel show scales in gravitational units $r_g$, while the bars in the right bottom corner show scales in physical units (pc) assuming M87 BH mass.}  \label{fig:compare_snapshots}
\end{figure*}

The detailed accretion and feedback properties of all runs are listed in \autoref{tab:run_details}. We also present temperature snapshots of the $a_*=0.9$ runs in \autoref{fig:compare_snapshots}, taken during their active jet phases. The relativistic jet propagates beyond the Bondi radius $R_B$, and for the largest Bondi radius, $R_B\approx 2\times 10^6\,r_g$, the jet can extend to kpc scales.

\section{Lognormal distribution and standard error calculation}\label{sec:lognormal}

If $X$ follows a lognormal distribution, then $\ln X$ is normally distributed: $\ln X \sim \mathcal{N}(\mu_{\ln X},\sigma_{\ln X})$. The mean of $X$ is related to the statistical properties of $\ln X$ as 
\begin{equation}\label{eq:mean_lognormal}
    \mu_X = \exp{(\mu_{\ln X} + \sigma_{\ln X}^2/2)}.
\end{equation}
The confidence interval of the mean $X$ is
\begin{equation}
    \text{CI}(\mu_X)=\exp{\left(\mu_{\ln X} + \frac{\sigma_{\ln X}^2}{2}\pm \sqrt{\frac{\sigma_{\ln X}^2}{N_{\rm eff}} + \frac{\sigma_{\ln X}^4}{2(N_{\rm eff} -1)}}\right)}
\end{equation}
for $68\%$ level confidence \citep{Zhou1997,Olsson2005}. Here, $N_{\rm eff}$ is the effective independent sample size. The confidence intervals for each run are shown as errorbars in \autoref{fig:compare_quantity_rB}. 

Since the time series data $\dot{M}(t)$ and $\eta(t)$ are temporarily correlated, we adopt the effective sample size $N_{\rm eff}$ instead of the total sample size $N$. We estimate the effective sample size $N_{\rm eff}$ by first taking the autocorrelation of the time series, and dividing $N$ by the autocorrelation length.

\bibliography{bridging_scales}{}
\bibliographystyle{aasjournal}

\end{CJK}
\end{document}